\documentclass[journal]{IEEEtran}
\usepackage{longtable,supertabular}
\usepackage{amsmath}
\usepackage{amssymb}
\usepackage{latexsym}
\usepackage{cite}

\ifCLASSINFOpdf
\else
\fi
\begin{document}
\title{New Explicit Good Linear Sum-Rank-Metric Codes}

\author{Hao Chen% <-this % stops a space
\thanks{Hao Chen is with the College of Information Science and Technology/Cyber Security, Jinan University, Guangzhou, Guangdong Province, 510632, China, haochen@jnu.edu.cn. The research of Hao Chen was supported by NSFC Grant 62032009.}% <-this % stops a space
\thanks{Manuscript received February 9, 2023; revised April 13, 2023.}}


\markboth{IEEE TRANSACTIONS ON INFORMATION THEORY,~Vol.~ , No.~ , June~2023}%
{Shell \MakeLowercase{\textit{et al.}}: Bare Demo of IEEEtran.cls for IEEE Journals}



\maketitle

\begin{abstract}
  Sum-rank-metric codes have wide applications in universal error correction, multishot network coding, space-time coding and the construction of partial-MDS codes for repair in distributed storage. Fundamental properties of sum-rank-metric codes have been studied and some explicit or probabilistic constructions of good sum-rank-metric codes have been proposed. In this paper we give three simple constructions of explicit linear sum-rank-metric codes. In finite length regime, numerous larger linear sum-rank-metric codes with the same minimum sum-rank distances as the previous constructed codes can be derived from our constructions. For example several better linear sum-rank-metric codes over ${\bf F}_q$ with small block sizes and the matrix size $2 \times 2$ are constructed for $q=2, 3, 4$ by applying our construction to the presently known best linear codes. Asymptotically our constructed sum-rank-metric codes are close to the Gilbert-Varshamov-like bound on sum-rank-metric codes for some parameters.  Finally we construct a linear MSRD code over an arbitrary finite field ${\bf F}_q$ with various square matrix sizes $n_1, n_2, \ldots, n_t$ satisfying $n_i \geq n_{i+1}^2+\cdots+n_t^2$ , $i=1, 2, \ldots, t-1$, for any given minimum sum-rank distance. There is no restriction on the block lengths $t$ and parameters $N=n_1+\cdots+n_t$ of these linear MSRD codes from the sizes of the fields ${\bf F}_q$.
\end{abstract}


% \fi
\IEEEpeerreviewmaketitle

\section{Introduction}
\IEEEPARstart{F}{or} a vector ${\bf a} \in {\bf F}_q^n$, the Hamming weight $wt_H({\bf a})$ of ${\bf a}$ is the number of non-zero coordinate positions. The Hamming distance $d_H({\bf a}, {\bf b})$ between two vectors ${\bf a}$ and ${\bf b}$ is defined as $wt_H({\bf a}-{\bf b})$. For a code ${\bf C} \subset {\bf F}_q^n$, its Hamming distance is the minimum of Hamming distances $d_H({\bf a}, {\bf b})$ between any two different codewords ${\bf a}$ and ${\bf b}$ in ${\bf C}$, $$d_H=\min_{{\bf a} \neq {\bf b}} \{d_H({\bf a}, {\bf b}): {\bf a}, {\bf b} \in {\bf C}\}.$$ It is well-known that the Hamming distance of a linear code ${\bf C}$ is the minimum Hamming weight of its non-zero codewords. The theory of Hamming metric error-correcting codes has been extensively studied and numerous constructions have been proposed, see e.g. \cite{HP}.   For a linear $[n, k, d_H]_q$ code, the Singleton bound asserts $d_H \leq n-k+1$. When equality holds, this code is called a maximal distance separable (MDS) code. The main conjecture of MDS codes claims that the length of an MDS code over ${\bf F}_q$ is at most $q+1$, except some special cases. In \cite{Ball} the main conjecture of MDS codes was proved for codes over prime fields.

In this paper the repetition code in the Hamming metric ${\bf C}=\{(c_1, \ldots, c_n):c_1=\cdots=c_n\}$ over some finite field will be used. For each nonzero codeword in this code, the Hamming weight is exactly $n$.

The rank-metric on the space ${\bf F}_q^{(m, n)}$ of size $m \times n$ matrices over ${\bf F}_q$ is defined by the ranks of matrices, $d_r(A,B)= rank(A-B)$. The minimum rank-distance of a code ${\bf C} \subset {\bf F}_q^{(m, n)}$  is $$d_r({\bf C})=\min_{A\neq B} \{d_r(A,B): A, B \in {\bf C} \}.$$  The rate of this code ${\bf C}$ is $rate({\bf C})=\frac{\log_q |{\bf C}|}{mn}$. For a code ${\bf C}$ in ${\bf F}_q^{(m, n)}$ with the minimum rank distance $d_r({\bf C}) \geq d$, it is well-known that the number of codewords in ${\bf C}$ is upper bounded by $q^{\max\{m,n\}(\min\{m,n\}-d+1)}$ , see \cite{Gabidulin}.  A code attaining this bound is called a maximal rank distance (MRD) code. The Gabidulin code $Gab(n, v) \subset {\bf F}_q^{(n, n)}$ is consisting of ${\bf F}_q$-linear mappings on ${\bf F}_q^n \cong {\bf F}_{q^n}$ defined by $q$-polynomials $a_0x+a_1x^q+\cdots+a_ix^{q^i}+\cdots+a_tx^{q^t}$, where $a_t,\ldots,a_0 \in {\bf F}_{q^n}$ are arbitrary elements in ${\bf F}_{q^n}$, see \cite{Gabidulin}. The rank-distance of the Gabidulin code is at least $n-t$ since there are at most $q^t$ roots in ${\bf F}_{q^n}$ for each such $q$-polynomial. There are  $q^{n(t+1)}$ such $q$-polynomials. Hence the size of the Gabidulin code is $q^{n(t+1)}$ and it is an MRD code.  Let $h$ be a non-negative integer and $\phi: {\bf F}_{q^k} \longrightarrow {\bf F}_{q^{k+h}}$ be a ${\bf F}_q$-linear embedding. Then $$a_t \phi(x)^{q^t}+a_{t-1}\phi(x)^{q^{t-1}}+\cdots+a_1\phi(x)^q+a_0\phi(x)$$ is a ${\bf F}_q$-linear mapping from ${\bf F}_{q^k}$ to ${\bf F}_{q^{k+h}}$, where $a_i \in {\bf F}_{q^{k+h}}$ for $i=0,1,\ldots,t$. It is clear that the dimension of the kernel of any such mapping is at most $t$. Then the rank-metric code consisting of all such linear mappings is an MRD code with rank distance $k-t$ and size $q^{(k+h)(t+1)}$ elements. MRD codes have been widely used in previous constructions of constant dimension subspace codes, see \cite{ES09}.  We refer to \cite{BNRS,Gorla,Bartz} for recent results on rank-metric codes and \cite{HTM2017,NHTRR2018} for recent results on MRD codes.

Sum-rank-metric codes have applications in multishot network coding, see \cite{MK19,NPS,NU,WZSS}, space-time coding, see \cite{SK}, and coding for distributed storage see \cite{CMST,MK,MP1}. For fundamental properties and constructions of sum-rank-metric codes, we refer to \cite{MK,MP1,BGR,BGR1,CGLGMP,OPB,MP21,MP22,MPK22}. Now we recall some basic concepts and results for sum-rank-metric codes in \cite{BGR}. Let $n_i \leq m_i$ be $2t$ positive integers satisfying $m_1 \geq m_2 \cdots \geq m_t$. Set $N=n_1+\cdots+n_t$.  Let $${\bf F}_q^{(n_1, m_1), \ldots,(n_t, m_t)}={\bf F}_q^{n_1 \times m_1} \bigoplus \cdots \bigoplus {\bf F}_q^{n_t \times m_t}$$ be the set of all ${\bf x}=({\bf x}_1,\ldots,{\bf x}_t)$, where ${\bf x}_i \in {\bf F}_q^{n_i \times m_i}$, $i=1,\ldots,t$, is a $n_i \times m_i$ matrix over ${\bf F}_q$. We call $n_i \times m_i$, $i=1, \ldots, t$,  matrix sizes of sum-rank-metric codes. Set $wt_{sr}({\bf x}_1, \ldots, {\bf x}_t)=rank({\bf x}_1)+\cdots+rank({\bf x}_t)$ and $$d_{sr}({\bf x},{\bf y})=wt_{sr}({\bf x}-{\bf y}),$$ for ${\bf x}, {\bf y} \in {\bf F}_q^{(n_1,m_1), \ldots,(n_t,m_t)}$. This is indeed a metric on ${\bf F}_q^{(n_1,m_1), \ldots,(n_t,m_t)}$. \\

{\bf Definition 1.1.} {\em A sum-rank-metric code ${\bf C} \subset {\bf F}_q^{(n_1,m_1), \ldots,(n_t,m_t)}$ is a subset of  the finite metric space ${\bf F}_q^{(n_1,m_1), \ldots,(n_t,m_t)}$. Its minimum sum-rank distance is defined by $$d_{sr}({\bf C})=\min_{{\bf x} \neq {\bf y}, {\bf x}, {\bf y} \in {\bf C}} d_{sr}({\bf x}, {\bf y}).$$  The code rate of ${\bf C}$ is $R_{sr}=\frac{log_q |{\bf C}|}{\Sigma_{i=1}^t n_im_i}$. The relative distance is $\delta_{sr}=\frac{d_{sr}}{N}$.}

The basic goal of sum-rank-metric coding is to construct good sum-rank-metric codes with large cardinalities and large minimum sum-rank distances. For some basic upper bounds on sizes of sum-rank-metric codes, we refer to \cite[Section III \& IV]{BGR}.

The following several special cases of parameters are important. For $t=1$, this is the rank-metric code case. For  $m_1=\cdots=m_t=m$ and $n_1=\cdots=n_t=n$, this is the $t$-sum-rank-metric code over ${\bf F}_{q^m}$ with the code length $N=nt$. For $m=n=1$, this is the Hamming metric error-correcting code case. Hence the sum-rank-metric is a generalization and combination of the Hamming metric and the rank-metric.

A Singleton-like bound for the sum-rank-metric was proposed in \cite{MK,BGR}. The general form Theorem III.2 in \cite{BGR} is as follows. The minimum sum-rank distance $d$ can be written uniquely as the form $d_{sr}=\Sigma_{i=1}^{j-1} n_i+\delta+1$ where $0 \leq \delta \leq n_j-1$, then $$|{\bf C}| \leq q^{\Sigma_{i=j}^t n_im_i-m_j\delta}.$$ The code attaining this bound is called a maximal sum-rank-metric distance (MSRD) code. When $m_1=\cdots=m_t=m$, this bound is of the form $$|{\bf C}| \leq q^{m(N-d_{sr}+1)}.$$  We call the difference $m(N-d_{sr}+1)-\log_q |{\bf C}|$ the {\em defect} of the sum-rank-metric code ${\bf C}$.

When $t \leq q-1$ and $N \leq (q-1)m$, MSRD codes attaining the above Singleton-like bound were constructed in \cite{MP1,MP22}. They are called linearized Reed-Solomon codes, we also refer to \cite{Neri} for the further results. When $t=q$, it was proved in \cite{BGR} Example VI.9, MSRD codes may not exist for some minimum sum-rank distance. In \cite{BGR} the maximal block lengths of MSRD codes are upper bounded in Theorem VI. 12. In several other cases, for example, when the minimum sum-metric distance is $2$ or $N$, MSRD codes exist for all parameters, they were constructed in Section VII of \cite{BGR}. In \cite{MP20} more linear MSRD codes with the same matrix size defined over smaller fields were constructed by extended Moore matrices.

One-weight sum-rank-metric codes were studied and constructed in \cite{MP191,NSZ21}. Generalized sum-rank weights were defined for sum-rank-metric codes via optimal anticodes in \cite{CGLGMP}. It was proved in \cite{CGLGMP} that the generalized sum-rank weights of an MSRD code is determined by its block size, matrix size, dimension and distance parameters, as that of the generalized Hamming weights of an MDS code, see \cite{Wei,Rav2016}. MSRD codes have applications in space-time coding, see e.g. \cite{SK}, maximally recoverable LRC codes and partial-MDS codes, see e.g. \cite{MK,CMST}.

Sum-rank BCH codes of the matrix size $n_1=\cdots=n_t=n$, $m_1=\cdots=m_t=m$ were proposed and studied in \cite{MP21} by the deep algebraic method. These sum-rank-metric codes are linear over ${\bf F}_{q^m}$.  There is a designed distance of such a sum-rank BCH code such that the minimum sum-rank distance is always greater than or equal to the designed distance.  On the other hand, the dimension of these sum-rank BCH codes is lower bounded in \cite{MP21} Theorem 9. Many sum-rank-metric codes with parameters $n=m=2$ and $q=2$ were constructed in Tables V, VI and VII of \cite{MP21}. It will be shown in Section III and the Appendix, many of our constructed sum-rank-metric codes are larger than these sum-rank BCH codes constructed in \cite{MP21} of the same minimum sum-rank distances.

The volume of radius $r$ in the sum-rank-metric is
\begin{align*}
  & vol(B_r({\bf F}_q^{(n_1,m_1), \ldots,(n_t,m_t)})) \\
  = & \Sigma_{s=0}^r \Sigma_{(s_0,\ldots,s_t): s_0+\cdots+s_t=s} \prod \displaystyle{n_i \choose s_i} \prod_{j=0}^{s_i-1}(q^{m_i}-q^j),
\end{align*}
we refer to Lemma III.5 in \cite{BGR}. In the case $n_1=\cdots=n_t=n_t$, $m_1=\cdots=m_t=m$, set $$f(z)=\Sigma_{i=0}^n \displaystyle{n \choose i} \prod_{j=0}^{i-1}(q^m-q^j)z^i,$$ and $$H_{sr}(\rho)=\frac{1}{mn}\min_{z \in (0,1]} log_q (\frac{f(z)}{z^{\rho}}),$$  where $\rho$ is a positive real number satisfying  $0<\rho<n$. Then from \cite{BGR} Theorem IV.9, when $n,m, \rho <n$ are fixed, $$\lim_{t \longrightarrow \infty} \frac{log_{q^{mn}} (vol(B_{\rho t}(({\bf F}_q^{(n,m), \ldots,(n,m)}))))}{t}=H_{sr}(\rho).$$ This is an entropy function for the sum-rank. From Lemma 2 in \cite{OPB} we have $$H_{sr}(\rho) \geq \frac{(m+n-\rho)\rho-\frac{1}{4}-log_q \gamma_q}{mn},$$ where $\gamma_q=\prod_{i=1}^{\infty}(1-q^{-i})^{-1}$, for example $\gamma_2 \approx3.463, \gamma_3 \approx1.785$ and $\gamma_4 \approx1.452.$ Non-list-decodability of Gabidulin codes and linearized Reed-Solomon codes as the sum-rank code generalization of Gabidulin codes were studied in\cite{WZ,PR}, which are closely related to the above entropy function for the sum-rank.

We now recall the sum-rank-metric Gilbert-Varshamov-like bound given in \cite{Puchinger,OPB} for the case $n_1=\cdots=n_t=n$ and $m_1=\cdots=m_t=m$.

{\bf Asymptotic Gilbert-Varshamov-like bound.} {\em For fixed positive integers $n$ and $m$, and positive integers $t, N=nt$, positive real number $R_{sr}$ and $\delta_{sr}=\frac{d}{N}$ and $2<d \leq N$ satisfying $$ R_{sr} \leq \delta_{sr}^2 \frac{n}{m}-\delta_{sr}(1+\frac{n}{m}+2\frac{n}{Nm})+1+\frac{1}{N}+\frac{n}{Nm}+\frac{n}{N^2m}$$ $$- \frac{\Sigma_{i=1}^{\delta_{sr} N-1} \log_q(1+\frac{t-1}{i})+\log_q(\delta_{sr} N-1)}{Nm}-\frac{\log_q (\gamma_q)}{nm},$$ there exists a linear sum-metric code of rate $R_{sr}$ and the relative minimum sum-rank distance at least $\delta_{sr}$. When $m=\xi n$ goes to the infinity and $m \in \omega(log_q(t))$, where $\xi$ is a fixed constant, then $$R_{sr} \sim \delta_{sr}^2-\delta_{sr}(1+\frac{1}{\xi})+1.$$}

It was proved in \cite{OPB} that random linear sum-rank-metric codes attain the Gilbert-Varshamov-like bound with high probability.

In this paper we give three constructions of linear sum-rank-metric codes. Our constructions are based by combining Hamming metric codes and $q$-polynomial representations of rank-metric codes. These constructions works for various block lengths and matrix sizes.  Then many explicit good linear sum-rank-metric codes are constructed from our second construction and the presently known best codes over ${\bf F}_4$ and ${\bf F}_9$ from \cite{codetable}. The linear sum-rank-metric codes constructed in this paper are good in the sense that our constructed codes are closer to the Singleton-like bound or are larger when compared to previous codes of the same minimum sum-rank distances. Asymptotically good linear sum-rank-metric codes which are close to the Gilbert-Varshamov like bound are also given. Sum-rank-metric codes constructed in this paper is clearly ${\bf F}_q$-linear, not always ${\bf F}_{q^m}$-linear as in the sense of \cite{MP1,MP21}.

There have been several constructions of sum-rank-metric MSRD codes of various matrix sizes, we refer to \cite{BGR,MP20,Bartz2}. In previous papers about constructions of sum-rank-metric codes \cite{MP1,MK19,MP21,BGR,BGR1,CGLGMP,Neri,NSZ21}, the main focus is on the matrix size case $n_1=n_2=\cdots=n_t, m_1=m_2=\cdots=m_t$. MSRD codes over a fixed field with an arbitrary block length and the matrix size $n_i \times m_i$, $n_i \neq m_i$, were constructed in \cite[Subsection 4.5]{MP20}. In Section V we construct new linear MSRD codes over an arbitrary finite field ${\bf F}_q$ with square matrix sizes $n_1 \times n_1, n_2 \times n_2, \ldots, n_t\times n_t$, where $n_1, n_2, \ldots, n_t$ are $t$ positive integers satisfying $n_i \geq n_{i+1}^2+\cdots+n_t^2$, for $i=1, 2, \ldots, t-1$,  and any given minimum sum-rank distance $d_{sr}$. There is no restriction on the length of the code from the size $q$ of the finite field.  Our result illustrates that the theory of sum-rank-metric codes with various square matrix sizes is basically different with the theory of sum-rank-metric codes with the same square matrix size. This is also quite different to the essence of the main conjecture on MDS codes in the Hamming metric. On the other hand, comparing with constructed non-trivial optimal LRC codes, quantum MDS codes and entanglement-assisted quantum MDS codes attaining the Singleton bound in \cite{GXY,CZJL,Pellikaan}, code lengths are bounded by some $O(q^2)$, the block lengths $t$ and the parameters $N=n_1+\cdots+n_t$ of MSRD codes with various square matrix sizes, have no relation with the field ${\bf F}_q$.

\section{Explicit constructions of sum-rank-metric codes}

In this section we give our first and second explicit constructions of linear sum-rank-metric codes. The matrix size is restricted to the case $n_1=\cdots=n_t=n=m_1=\cdots=m_t$. The construction can be generalized to the matrix size case $n_1=n_2=\cdots=n_t$, $n_i \leq m_i$, $i=1,2, \ldots, t$ directly.

\subsection{Construction 1: One Hamming metric code over the large field}

The Gabidulin code $Gab(n, v)$ in ${\bf F}_q^{(n, n)}$ of the minimum rank distance $n-v$ and cardinality $q^{n(v+1)}$ is identified with the set of all $q$-polynomials $a_0x+a_1 x^q+\cdots+a_v x^{q^v}$, where $a_0, \ldots, a_v$ are arbitrary elements in ${\bf F}_{q^n}$. Hence $(a_0, a_1, \ldots, a_v)$ can be considered as elements in the finite field ${\bf F}_{q^{n(v+1)}}$. Let ${\bf C}$ be a linear $[t, w, d]_{q^{n(v+1)}}$ code over this finite field ${\bf F}_{q^{n(v+1)}}$, with the dimension $w$ and minimum Hamming distance $d$. Then we have a linear sum-rank-metric code $SR({\bf C})$ consisting of all $({\bf c}^1, \cdots, {\bf c}^t)$ where ${\bf c}^i=(c_0^i, c_1^i, \ldots, c_v^i)$, $c_j^i \in {\bf F}_{q^n}$,  are considered as a $q$-polynomial $c_0^ix+c_1^i x^q+\cdots+c_v^i x^{q^v}$ associated with the codeword in $Gab(n, v)$. This code $SR({\bf C})$ is ${\bf F}_{q^n}$-linear.

{\bf Remark 2.1.} From the coordinate form of the code $SR({\bf C})$, the above construction is similar to the construction for convolutional codes in \cite{NPS}.

The following result is obvious.

{\bf Proposition 2.1.} {\em The dimension over ${\bf F}_q$ of the above linear sum-rank-metric code $SR({\bf C})$ is $wn(v+1)$ and the minimum rank-sum distance of $SR({\bf C})$ is at least $d(n-v)$.}

Hence the code rate of $SR({\bf C})$ is $R_{sr}(SR({\bf C}))=\frac{wn(v+1)}{tn^2}=\frac{w(v+1)}{tn}=R({\bf C}) \cdot \frac{v+1}{n}$ and the relative minimum sum-rank distance is at least $\frac{d(n-v)}{tn}=\delta({\bf C}) \cdot \frac{n-v}{n}$, where $R({\bf C})$ and $\delta({\bf C})$ are the rate and  the relative minimum Hamming distance of the code ${\bf C}$. Therefore for fixed $n$ and $v$, asymptotically good linear sum-rank-metric codes with positive rate and positive relative minimum sum-rank distance can be constructed from asymptotically good linear codes in the Hamming metric. Assume that $n$ is even, $v=n-2$,  and by the using of algebraic geometry code sequence satisfying $R+\delta \geq 1-\frac{1}{q^{\frac{n(v+1)}{2}}-1}$, see \cite{TV}, we can get asymptotically good sum-rank-metric code sequence with the rate $$R_{sr}=\frac{n-1}{n} \cdot R$$ and the relative minimum sum-rank distance at least $$\delta_{sr} \geq \frac{2}{n} \left(1-R-\frac{1}{q^{\frac{n(v+1)}{2}}-1}\right).$$ Then we have $$R_{sr}+2\delta_{sr}-\delta_{sr}^2 \geq \frac{4}{n}+\frac{n-5}{n}R-\frac{4}{n^2}(1-R)^2.$$ When $n$ is large and $R$ is close to $1$, the rate and the relative minimum sum-rank distance is close to the Gilbert-Varshamov-like bound $$R_{sr} \sim \delta_{sr}-2\delta_{sr}+1,$$ in \cite{OPB}, or see the bound cited in Section I.

{\bf Remark 2.2.} Comparing with the previous probabilistic construction of sum-rank-metric codes attaining the Gilbert-Varshamov-like bound in \cite{OPB}, the codes in the above construction is given with the help of algebraic geometry codes attaining the Tsfasman-Vl\'{a}dut-Zink bound. Then these sum-rank-metric codes can be constructed by a low-complexity polynomial-time algorithm. We refer to \cite{SAKSD} for a low-complexity polynomial-time algorithm constructing algebraic geometry codes attaining the Tsfasman-Vl\'{a}dut-Zink bound.

\subsection{Construction 2: Several Hamming metric codes over the small field}

A modification of the construction 1 makes the resulted sum-rank-metric codes larger. The first linear $[t, k_0, w_0]_{q^n}$ code ${\bf C}_0 \subset {\bf F}_{q^n}^t$ corresponds to $a_0^i$ in the $q$-polynomials $a_0^ix+a_1^i x^q+\cdots+a_v^ix^{q^v}$ in the $i$-th copy of Gabidulin code $Gab(n, v)$ at the $i$-th block position, for $i=1, 2, \ldots, t$. The second linear $[t, k_1, w_1]_{q^n}$ code ${\bf C}_1 \subset {\bf F}_{q^n}^t$ corresponds to $a_1^i$ in the $q$-polynomials $a_0^ix+a_1^i x^q+\cdots+a_v^ix^{q^v}$ in the $i$-th copy of Gabidulin code $Gab(n, v)$ at the $i$-th block position, $i=1, 2, \ldots, t$, ......, The $(v+1)$-th linear $[t, k_v, w_v]_{q^n}$ code ${\bf C}_v$ corresponds to $a_v^i$ in the $q$-polynomials $a_0^ix+a_1^i x^q+\cdots+a_v^ix^{q^v}$ in the $(v+1)$-th copy of Gabidulin code $Gab(n, v)$ at the $i$-the block position, $i=1, 2, \ldots, t$. The sum-rank-metric code $SR({\bf C}_0, \ldots, {\bf C}_v)$ is consisting of $q$-polynomials as follows,
\begin{align*}
  & SR({\bf C}_0, ..., {\bf C}_v) =\{(a_0^1x+a_1^1x^q+\cdots+a_v^1x^{q^v}, \ldots, \\
  & a_0^tx+a_1^tx^q+\cdots+a_v^t x^{q^v}):
   (a_0^1, \ldots, a_0^t) \in {\bf C}_0, \ldots, \\
  & (a_v^1, \ldots, a_v^t) \in {\bf C}_v\}.
\end{align*}
This is a linear (over ${\bf F}_q$, not ${\bf F}_{q^n}$) sum-rank-metric code with the minimum sum-rank distance at least $\min \{w_0n, w_1(n-1), \ldots, w_v(n-v)\}$. It is easy to verify the linear independence, therefore the dimension is $$\dim_{{\bf F}_q}(SR({\bf C}_0, \ldots, {\bf C}_v))=n(k_0+\cdots+k_v).$$

{\bf Theorem 2.1.} {\em Let ${\bf C}_i \subset {\bf F}_{q^n}^t$ be a linear $[t, k_i, w_i]_{q^n}$ code, for $i=0, 1, \ldots, v$. Then $SR({\bf C}_0, \ldots, {\bf C}_v)$ is a block length $t$ and matrix size $n \times n$ linear sum-rank-metric code over ${\bf F}_q$ of the dimension $n(k_0+\cdots+k_v)$. The minimum sum-rank distance of $SR({\bf C}_0, \ldots, {\bf C}_v)$ is at least $\min \{w_0n, w_1(n-1), \ldots, w_v(n-v)\}$.}

{\bf Proof.} The dimension can be calculated directly. The lower bound on the minimum sum-rank distance is from the formation of $q$-polynomials in this code.

{\bf Remark 2.3.} The main difference between the construction 2 and the construction 1 is as follows. When several codes over ${\bf F}_{q^n}$ are used in Theorem 2.1 to construct a linear sum-rank-metric code, $w_0, w_1, \ldots, w_v$ satisfy $w_0n=w_1(n-1)=\cdots=w_v(n-v)$, or these $v+1$ numbers $w_0n, w_1(n-1), \ldots, w_v(n-v)$ are close. Therefore the dimensions of the first several codes ${\bf C}_0, \ldots, {\bf C}_i$ can be larger, since their minimum Hamming distances are smaller. This property makes the sum-rank-metric codes in the construction 2 larger.

\subsection{Sum-rank-metric codes from Hamming metric BCH codes}

From Theorem 2.1, it is natural to use several Hamming metric BCH codes over ${\bf F}_{q^n}$ of length $q^{un}-1$ to construct sum-rank-metric codes of the block length $q^{un}-1$. Based on some previous calculations of dimensions of BCH codes in \cite{LDL,ZSK}, we get the lower bound on dimensions of these sum-rank-metric codes. It is interesting to notice that the parameters of these linear sum-rank-metric codes from Hamming metric BCH codes can be compared with the parameters of these sum-rank BCH codes developed by the deep algebraic method in \cite{MP21}. As shown in some examples below, some linear sum-rank-metric codes from BCH codes applied in Theorem 2.1 have almost the same dimensions as the smaller codes in \cite{MP21}. However the construction in Theorem 2.1 is direct and simple and can be applied to arbitrary Hamming metric codes, not only restricted to BCH codes.

{\bf Theorem 2.2.} {\em Let $q$ be a prime power and $u \geq 4$ be an even positive integer, $u_0, \ldots, u_v$, $v \leq n-1$, be positive integers satisfying $2 \leq u_i \leq q^{\frac{un}{2}}+1$, $i=0, 1, \ldots, v$. Then we have a block size $q^{un}-1$ and matrix size $n \times n$ linear sum-rank-metric code ${\bf C}$. The dimension of this linear sum-rank-metric code is at least $\dim_{{\bf F}_q}({\bf C})=n(\Sigma_{i=0}^v(q^{un}-1-u(u_i-1-\lfloor\frac{u_i-1}{q^n} \rfloor)))$, and the minimum sum-rank distance is at least $\min\{u_0n, u_1(n-1), \ldots, u_v(n-v)\}$.}

{\bf Proof.} We take primitive BCH codes over ${\bf F}_{q^n}$ of length $q^{nu}-1$ and the designed distance $u_0, u_1, \ldots, u_v$. The conclusion in the case 1 of Theorem 17 in \cite{LDL} asserts that there is a BCH code ${\bf C}_{q^{un}-1, q^n, u_i, 1}$ of the designed distance $u_i$ and the dimension $q^{un}-1-u(u_i-1-\lfloor \frac{u_i-1}{q^n}\rfloor)$. We take $v \leq n-1$ such BCH codes over ${\bf F}_{q^n}$ as codes ${\bf C}_0, \ldots, {\bf C}_v$ in Theorem 2.1. The conclusion follows immediately.

However to compare with the ${\bf F}_{q^m}$-linear sum-rank BCH codes constructed in \cite{MP21}, we can use primitive BCH codes over ${\bf F}_q$ of length $q^u-1$ and designed distances $u_0, \ldots, u_v$. These codes can be considered as linear codes over ${\bf F}_{q^n}$. Then we have the following sum-rank-metric codes as follows from Theorem 17 in \cite{LDL} about BCH codes over ${\bf F}_q$.

{\bf Corollary 2.1.} {\em Let $q$ be a prime power and $u \geq 4$ be an even positive integer, $u_0, \ldots, u_v$, $v \leq n-1$  be positive integers satisfying $2 \leq u_i \leq q^{\frac{u}{2}}+1$, $i=0, 1, \ldots, v$. Then we have a block length $q^u-1$ and matrix size $n \times n$ linear sum-rank-metric code ${\bf C}$ of the dimension $\dim_{{\bf F}_q}({\bf C})=n(\Sigma_{i=0}^v(q^{u}-1-u(u_i-1-\lfloor\frac{u_i-1}{q} \rfloor)))$. The minimum sum-rank distance of this sum-rank-metric code is at least $\min\{u_0n, u_1(n-1), \ldots, u_v(n-v)\}$.}

For $q=2$, $n=2$, $u=6$, $u_0=2$, $u_1=4$, this is a sum-rank-metric code of the block size $63$, with the dimension $\dim_{{\bf F}_2}=2((63-6)+(63-6 \cdot 2))=2 \cdot 108$, and minimum sum-rank distance $4$. In Table VI, page 5166 of \cite{MP21} there are two linear sum-rank-metric codes of the block size $63$ with minimum sum-rank distance $4$, the smaller code has the dimension lower bounded by $2\cdot 108$ and the larger code has the dimension lower bounded by $2 \cdot 112$. Hence our code have the same parameters as the smaller one constructed in \cite{MP21} by the deep algebraic method. For $q=2$, $n=2$, $u=6$, $u_0=3$, $u_1=6$, this is a sum-rank-metric code of the block size $63$, with the dimension $\dim_{{\bf F}_2}=2((63-6)+(63-6 \cdot 3))=2 \cdot 102$, and minimum sum-rank distance $6$. For $q=2$, $n=2$, $u=6$, $u_0=5$, $u_1=10$, this is a linear sum-rank-metric code of the block size $63$, with the dimension $\dim_{{\bf F}_2}({\bf C})=2((63-12)+(63-6 \cdot 5))=2 \cdot 84$, and minimum sum-rank distance $10$. In Table VI, page 5166 of \cite{MP21}, these two codes have the same parameters as the smaller codes constructed in \cite{MP21}.

{\bf Theorem 2.3.} {\em Let $q$ be a prime power and $u \geq 5$ be an odd positive integer, $u_0, \ldots, u_v$, $v\leq n-1$, be positive integers satisfying $2 \leq u_i \leq q^{\frac{(u+1)n}{2}}+1$, $i=0, 1, \ldots, v$. Then we have a block length $q^{un}-1$ and matrix size $n \times n$ linear sum-rank-metric code ${\bf C}$ of the dimension $\dim_{{\bf F}_q}({\bf C})=n(\Sigma_{i=0}^v(q^{un}-1-u(u_i-1-\lfloor\frac{u_i-1}{q^n} \rfloor)))$. The minimum sum-rank distance of this sum-rank-metric code is at least $\min\{u_0n, u_1(n-1), \ldots, u_v(n-v)\}$.}

{\bf Proof.} We take primitive BCH codes over ${\bf F}_{q^n}$ of length $q^{nu}-1$ and designed distance $u_0, u_1, \ldots, u_v$. The conclusion in the case 1 of Theorem 18 in \cite{LDL} asserts that there is a BCH code over ${\bf F}_{q^n}$ with the designed distance $u_i$ and the dimension $q^{un}-1-u(u_i-1-\lfloor \frac{u_i-1}{q^n}\rfloor)$. Then the conclusion follows from the construction Theorem 2.1 from $v+1$ such BCH codes over ${\bf F}_{q^n}$.

Form the conclusion in the case 1 of Theorem 18 in \cite{LDL} about BCH codes over ${\bf F}_q$, we get the following result.

{\bf Corollary 2.2.} {\em Let $q$ be a prime power and $u \geq 5$ be an odd positive integer, $u_0, \ldots, u_v$ be positive integers satisfying $2 \leq u_i \leq q^{\frac{u+1}{2}}+1$, $i=0, 1, \ldots, v$. Then we have a block length $q^u-1$ and matrix size $n \times n$ linear sum-rank-metric code ${\bf C}$ of the dimension $\dim_{{\bf F}_q}({\bf C})=n(\Sigma_{i=0}^v(q^{u}-1-u(u_i-1-\lfloor\frac{u_i-1}{q} \rfloor)))$. The minimum sum-rank distance of this sum-rank-metric code is at least $\min\{u_0n, u_1(n-1), \ldots, u_v(n-v)\}$.}

For $q=2$, $n=2$, $u=5$, the block length is $31$, we can construct a sum-rank-metric code with the dimension $\dim_{{\bf F}_2}=2((31-5)+(31-5 \cdot 3))=2 \cdot 42$, and minimum sum-rank distance $6$, a sum-rank-metric code with the dimension $\dim_{{\bf F}_2}=2((31-10)+(31-5 \cdot 4))=2 \cdot 32$, and  minimum sum-rank distance $8$, and a sum-rank-metric code with the dimension $\dim_{{\bf F}_2}({\bf C})=2((31-10)+(31-25))=2 \cdot 27$, and minimum sum-rank distance $10$.  In Table V, page 5165 of \cite{MP21} for each pair of above dimensions and minimum sum-rank distances, there are two linear sum-rank-metric codes, the smaller one has the same dimension and the minimum sum-rank distance as our code.

From the result in \cite{ZSK}, we have the following linear sum-rank-metric codes of the matrix size $n \times n$.

{\bf Theorem 2.4. } {\em Let $q$ be a prime power,  $\lambda$ and $n$ be two positive integers satisfying $\lambda | q^n-1$, $u$ be an odd positive integer, $u_0, \ldots, u_v$, $v\leq n-1$ be $v+1$ positive integers satisfying $1 \leq u_i-1 \leq \frac{q^{\frac{n(u+1)}{2}}-1}{\lambda}$, $i=0, 1, \ldots, v$.  Then we have a block length $\frac{q^{nu}-1}{\lambda}$ and matrix size $n \times n$ linear sum-rank-metric code ${\bf C}$. The dimension of this sum-rank-metric code is $\dim_{{\bf F}_q} ({\bf C})=n(\Sigma_{i=0}^v(\frac{q^{un}-1}{\lambda}-u\lceil (u_i-1)(1-\frac{1}{q^n})\rceil))$ and the minimum sum-rank distance of ${\bf C}$ is at least $\min \{u_0n, u_1(n-1), \ldots, u_v(n-v)\}$.}

{\bf Proof.} Theorem 1 in page 4701 of \cite{ZSK} asserts that there is a BCH code over ${\bf F}_{q^n}$ with designed distance $u_i$ and dimension $\frac{q^{un}-1}{\lambda}-u \lceil (u_i-1)(1-q^{-n})\rceil$. Then applying Theorem 2.1 to $v+1$ such BCH codes we get the conclusion.

For $q=2$, $n=2$, $\lambda=3$, a block length $341$ binary linear sum-rank-metric code of the dimension $607$ can be constructed from Theorem 3.3. The minimum sum-rank distance is at least $14$. There is no such sum-rank-metric code of the block length 341 constructed in \cite{MP21}. Notice that all previous results about BCH codes in the references of \cite{LDL,ZSK} of various lengths can be applied in our construction Theorem 2.1 to get linear sum-rank-metric codes. It is obvious that our construction Theorem 2.1 is simpler than the deep algebraic method in \cite{MP21}.

{\bf Remark 2.4.} BCH codes are not the presently known best linear codes for many parameters when compared to table \cite{codetable}. The above linear sum-rank-metric codes constructed from BCH codes can be improved significantly by applying codes \cite{codetable} in our construction 2. These linear sum-rank-metric codes from codes \cite{codetable} applied in the construction 2 will be give in Section III and the Appendix. We can see that many of them are larger than the linear sum-rank-metric codes constructed in \cite{MP21}.

\subsection{Sum-rank-metric codes from algebraic geometry codes}

Since we can use arbitrary error-correcting codes $({\bf C}_0, \ldots, {\bf C}_v)$ in the Hamming metric in our construction 2, it is natural to use the Reed-Solomon code and its generalization algebraic geometry codes to construct good linear sum-rank-metric codes. The advantage of this construction is the flexibility of block lengths of sum-rank-metric codes.

Let ${\bf F}_q$ be an arbitrary finite field, $P_1,\ldots,P_n$ be $n \leq q$ elements in ${\bf F}_q$. The Reed-Solomon code  $RS(n,k)$ is defined by $$RS(n,k)=\{(f(P_1),\ldots,f(P_n)): f \in {\bf F}_q[x],\deg(f) \leq k-1\}.$$ This is an MDS  $[n,k,n-k+1]_q$ linear code attaining the Singleton bound $d_H \leq n-k+1$, since  a degree $\deg(f) \leq k-1$ nonzero polynomial has at most $k-1$ roots.

{\bf Theorem 2.5.} {\em Let $q$ be a prime power, $1 \leq t \leq q^2$  and $1\leq k <t$ be two positive integers satisfying $t-k$ is odd. Then we have a block length $t$ and matrix size $2 \times 2$ linear sum-rank-metric code ${\bf C}$. The dimension of ${\bf C}$ is $\dim_{{\bf F}_q}({\bf C})=t+3k+1$ and the minimum sum-rank distance of this code ${\bf C}$ is at least $t-k+1$.}

{\bf Proof.} Applying Theorem 2.1 to the Reed-Solomon $[t, k, t-k+1]_{q^2}$ code and the Reed-Solomon $[t, \frac{t+k+1}{2}, \frac{t-k+1}{2}]_{q^2}$ code, we get the linear sum-rank-metric code.

Notice that linearized Reed-Solomon codes (MSRD) in \cite{MP1} can be constructed only if $t \leq q-1$. In Theorem 4.1, $t \leq q^2$, the defect is $2(2t-(t-k+1)+1)-(t+3k+1)=t-k-1$.

Reed-Solomon codes can be generalized to algebraic geometry codes, we refer to \cite[Chapter 13]{HP}. Let ${\bf X}$ be an absolutely irreducible  non-singular genus $g$ curve defined over ${\bf F}_q$ with $n$ rational points. Then a linear algebraic geometry $[n,m-g+1, \geq n-m]_q$ code can be constructed, where $m$ is the degree of the rational divisor satisfying $$2g-2 <m <n.$$ Reed-Solomon codes are just algebraic geometry codes over the genus $0$ curve. One achievement of the theory of algebraic geometry codes is the sequence of algebraic-geometric codes over ${\bf F}_{q^2}$ satisfying the Tsfasman-Vl\'{a}dut-Zink bound $$R+\delta \geq 1-\frac{1}{q-1},$$ which is exceeding the Gilbert-Varshamov bound when $q \geq 7$.  We refer to \cite{TV} for the detail.

We restrict to the case $n_1=\cdots=n_t=m_1=\cdots=n$, then length $t$ algebraic-geometric codes over ${\bf F}_{q^n}$ are used to construct sum-rank-metric codes as follows.

{\bf Theorem 2.6.} {\em Let ${\bf X}$ be an absolutely irreducible  non-singular genus $g$ curve defined over ${\bf F}_{q^n}$ with $N+1$ rational points (over ${\bf F}_{q^n}$). Let $u_0, \ldots, u_v$ be $v$ positive integers satisfying $2g-2<u_i<N$, $i=0, 1, \ldots, v$. Then we have a block length $N$ and matrix size $n \times n$ linear rank-sum-metric code ${\bf C}$ over ${\bf F}_q$. The dimension of this code ${\bf C}$ is $\dim_{{\bf F}_q}({\bf C})=n(\Sigma_{i=0}^v (u_i-g+1))$ and the minimum sum-rank distance of ${\bf C}$ is at least $\min \{(N-u_0)n, (N-u_1)(n-1), \ldots, (N-u_v)(n-v)\}$.}

From the curves over ${\bf F}_4$ in \cite{Geer} some sum-rank-metric codes over ${\bf F}_2$ of the matrix size $2 \times 2$ can be constructed as follows. For example we take a genus $50$ curve over ${\bf F}_4$ with $91$ rational points, then a block length $90$ and matrix size $2 \time 2$ linear sum-rank-metric code of the dimension $140$ is constructed. The minimum sum-rank distance is at least $8$. From algebraic geometry codes over  a genus $12$ curve over ${\bf F}_8$ with $49$ rational points, see \cite{Geer}, a block lenght $48$ and matrix size $2 \times 2$ linear sum-rank-metric code of the dimension $89$ is constructed. The minimum sum-rank distance is at least $12$.

\section{Linear sum-rank-metric codes from the presently known best codes}

Thanks to the table of the presently known best linear codes in \cite{codetable}, numerous good small block length linear sum-rank-metric codes over ${\bf F}_{q}$ of the matrix size $2 \times 2$, for $q=2, 3, 4$, can be constructed explicitly from Theorem 2.1. We give tables to list some linear sum-rank-metric codes constructed from Theorem 2.1 and the presently known best Hamming metric codes of \cite{codetable} in this section and the Appendix. Many of constructed sum-rank-metric codes have larger dimensions when compared with codes constructed in \cite{MP21}. It is clear that many of our constructed codes from Theorem 2.1 are close to the Singleton-like bound. For example when the block length is $7$ we have the following table of linear sum-rank-metric codes.

\begin{table}[h!]
\caption{Block length $t=7$, $n=m=2$.}
\label{tab:A-q-5-3}
\begin{center}
\begin{tabular}{|l|l|l|l|}
\hline
$d_{sr}$&dimension &Table III, \cite{MP21}&Singleton \\ \hline
$2$& $2 \cdot 13$  & $2 \cdot 12$&$2 \cdot 13$ \\ \hline
$3$ & $2 \cdot 10$  &$2 \cdot 9$&$2 \cdot 12$  \\ \hline
$4$ &$2 \cdot 9$  &$ 2 \cdot 6$&$2 \cdot 11$ \\ \hline
$5$ &$2 \cdot 6$ & $2 \cdot 5$&$2\cdot 10$ \\ \hline
$6$ &$ 2\cdot 5$  & $2 \cdot 5$ &$2 \cdot 9$\\ \hline
$7$ &$ 2 \cdot 4$ & $2 \cdot 2$&$2 \cdot 8$ \\ \hline
\end{tabular}
\end{center}
\end{table}

The following table lists the block length $31$, the matrix size $2 \times 2$, binary linear sum-rank-metric codes. Comparing with Table V in \cite{MP21} our codes are larger. Many of our codes from Theorem 2.1 are close to the the Singleton-like bound.

\begin{table}
\caption{Block length $t=31$, $n=m=2$.}
\label{tab:A-q-5-3}
\begin{center}
\begin{tabular}{|l|l|l|l|}
\hline
$d_{sr}$&dimension &Table V, \cite{MP21}&Singleton \\ \hline
$4$& $2 \cdot 55$  &$ 2 \cdot 50$ & $2 \cdot 59$ \\ \hline
$5$ & $2 \cdot 51$  &$2 \cdot 47$ & $2 \cdot 58$ \\ \hline
$6$ &$2 \cdot 50$  &$2 \cdot 45$ & $2 \cdot 57$ \\ \hline
$7$ &$2 \cdot 46$ & $2 \cdot 40$ & $2 \cdot 56$ \\ \hline
$8$ &$ 2\cdot 45$  & $2 \cdot 35$ & $2 \cdot 55$ \\ \hline
$9$ &$ 2 \cdot 41$ & none & $2 \cdot 54$ \\ \hline
$10$ &$2 \cdot 40$ &$2 \cdot 30$ & $2 \cdot 53$ \\ \hline
$11$ &$2 \cdot 38$ &none & $2 \cdot 52$ \\ \hline
$12$ &$2 \cdot 38$ &$2 \cdot 25$  & $2 \cdot 51$   \\ \hline
$13$ &$ 2 \cdot 33$ & none  & $2 \cdot 50$ \\ \hline
$14$ &$2 \cdot 31$ &$2 \cdot 20$ & $2 \cdot 49$ \\ \hline
$15$ &$2 \cdot 29$  & none  & $2 \cdot 48$ \\ \hline
$16$ &$2 \cdot 28$  & none  & $2 \cdot 47$ \\ \hline
$17$ &$2 \cdot 25$  & none  & $2 \cdot 46$ \\ \hline
$18$& $2 \cdot 24$  & $2 \cdot 7$ & $2 \cdot 45$ \\ \hline
$19$ & $2 \cdot 22$  &none & $2 \cdot 44$ \\ \hline
$20$ &$2 \cdot 21$  &none & $2 \cdot 43$ \\ \hline
$21$ &$2 \cdot 19$ & none & $2 \cdot 42$ \\ \hline
$22$ &$ 2\cdot 19$  & $2 \cdot 7$ & $2 \cdot 41$ \\ \hline
$23$ &$ 2 \cdot 18$ & none & $2 \cdot 40$ \\ \hline
$24$ &$2 \cdot 17$ &none & $2 \cdot 39$ \\ \hline
$25$ &$2 \cdot 13$ &none  & $2 \cdot 38$ \\ \hline
$26$ &$2 \cdot 13$ &$ 2\cdot 2$  & $2 \cdot 37$   \\ \hline
$27$ &$ 2 \cdot 12$ & none  & $2 \cdot 36$ \\ \hline
$28$ &$2 \cdot 12$ &none & $2 \cdot 35$ \\ \hline
$29$ &$2 \cdot 11$  & none  & $2 \cdot 34$ \\ \hline
$30$ &$2 \cdot 11$  & $2 \cdot 2$  & $2 \cdot 33$ \\ \hline
\end{tabular}
\end{center}
\end{table}

More good linear sum-rank-metric codes from the codes in \cite{codetable} applied in the construction 2 are given in the Appendix.

{\bf Remark 3.1.} However it should be indicated that the true dimensions and the true minimum sum-rank distances were not calculated explicitly in \cite{MP21}. The entries from \cite{MP21} in Table 1, 2 and the tables in the Appendix are lower bounds of dimensions and distances.

\section{Asymptotically good sum-rank-metric codes close to the GV-like bound}

In this section let $q$ be a fixed prime power and the square matrix size parameter $n$ will be a sufficiently large even positive integer. Asymptotically good sequences of ${\bf F}_q$-linear sum-rank-metric codes over ${\bf F}_q$ close to the Gilbert-Varshamov-like bound in \cite{OPB} are presented. In Theorem 2.1, let $v \leq n-1$ be a fixed positive integer, we need a linear $[n(n-1)\cdots (n-v)w_j, k_i, \frac{n(n-1) \cdots (n-v)w_j}{n-i}]_{q^n}$ code for $i=0, 1, \ldots, v$, where $w_j$ be a sequence of positive integers going to the infinity.  From Theorem 2.6 about sum-rank-metric codes from  algebraic geometry codes over ${\bf F}_{q^n}$, we take a family of algebraic curves $\{C_j\}_{j=1, 2, \ldots}$ over ${\bf F}_{q^n}$, with the genus $\{g_j\}_{i=1, 2, \ldots}$ and $\{N(C_i)\}_{i=1, 2, \ldots}$ rational points satisfying $N(C_i)-1\geq n(n-1) \cdot (n-v)w_j$ rational points, see \cite{TV}. Then we have $$\frac{k_i}{n(n-1) \cdots (n-v)w_j} \geq 1- \frac{1}{n-i}-\frac{1}{q^{\frac{n}{2}}-1},$$ from the Tsfasman-Vl\'{a}dut-Zink bound. Therefore the minimum sum-rank distance is at least $n(n-1) \cdots (n-v)w_j$. The relative minimum sum-rank distance is at least $$\delta_{sr}=\frac{1}{n}.$$ The dimension over ${\bf F}_q$ is $n\Sigma_{i=0}^v k_i$ and the code rate of this sum-rank-metric code is at least $$R_{sr} \geq \frac{1}{n}\left(v+1-\Sigma_{i=0}^v \frac{1}{n-i}-\frac{v+1}{q^{\frac{n}{2}}-1}\right).$$ Hence $$R_{sr}+2\delta_{sr}-\delta_{sr}^2 \geq \frac{1}{n}\left(v+3-\Sigma_{i=0}^v \frac{1}{n-i}-\frac{1}{q^{n/2}-1}-\frac{1}{n}\right).$$

When $v=n-2$, $R$ and $\delta$ satisfy the following  $$R_{sr}+2\delta_{sr} \geq 1+\frac{1}{n}-\frac{1}{n}\left(\frac{1}{n}+\frac{1}{n-1}+\cdots+\frac{1}{2}\right)-\frac{n-1}{q^{\frac{n}{2}}-1}.$$

{\bf Theorem 4.1.} {\em For any fixed square prime power $q$ and positive integers $n$ and $v$ satisfying $v\leq n-1$, we have a sequence of ${\bf F}_q$-linear sum-rank-metric codes with positive code rate $R_{sr}$ and positive relative minimum sum-rank distance $\delta_{sr}$, satisfying $$R_{sr}+2\delta_{sr}-\delta_{sr}^2 \geq \frac{1}{n}\left(v+3-\Sigma_{i=0}^v \frac{1}{n-i}-\frac{1}{q^{n/2}-1}-\frac{1}{n}\right).$$}

 From the asymptotical Gilbert-Varshamov-like bound on sum-rank-metric codes in \cite{OPB}, or see Section I, we have the following result.

{\bf Corollary 4.1.} {\em When $q$ is fixed and $n$ is sufficiently large, sequences of asymptotically good ${\bf F}_q$-linear sum-rank-metric codes close to the Gilbert-Varshamov-like bound can be constructed.}

{\bf Proof.} We consider the harmonic series $$\lim_{n \longrightarrow \infty}\left(1+\frac{1}{2}+\cdots+\frac{1}{n}-ln n\right)=\gamma,$$ where $\gamma=0.577215...$ is the Euler-Mascheroni constant. Then we have $$R_{sr}+2\delta_{sr}-\delta_{sr}^2 \geq 1-\frac{1}{q^{n/2}-1}-\frac{lnn}{n}$$ from Theorem 4.1. It is clear that $\lim_{n \longrightarrow \infty}\frac{lnn}{n}=0$ and the above sequence of linear sum-rank-matric codes has their $R_{sr}$ and $\delta_{sr}$ close to the GV-like bound $$R_{sr} \sim \delta_{sr}^2-2\delta_{sr}+1,$$ when $n$ is sufficiently large.

{\bf Remark 4.1.} The only non-explicitness of these codes in Theorem 4.1 is from the fact that the algebraic geometry code sequence achieving the Tsfasman-Vl\'{a}dut-Zink bound has not been constructed explicitly. However these codes in Corollary 4.1 can be constructed by a low-complexity polynomial time algorithm, see \cite{SAKSD}. It seems that both construction 1 and 2 are not sufficient to construct a sequence of sum-rank-metric codes achieving or exceeding the GV-like bound.

\section{Linear MSRD codes}

\subsection{Block size two MSRD codes}

In this Section we first give linear MSRD codes of block size $t=2$ for the convenience of understanding. First of all the matrix space ${\bf M}_{n \times n}({\bf F}_q)$ is identified with all $q$-polynomials $a_0 +a_1x^q+\cdots+a_{n-1}x^{q^{n-1}}$, where $a_0, a_1, \ldots, a_{n-1} \in {\bf F}_{q^n}$.

{\bf Theorem  5.1.} {\em Let $n_1$ and $n_2$ be two positive integers satisfying $n_1 \geq n_2^2$. Then a linear MSRD code ${\bf C}$ with the block size $2$, matrix sizes $n_1 \times n_1, n_2\times n_2$ over an arbitrary field ${\bf F}_q$ and any given minimum sum-rank distance can be constructed explicitly. }

{\bf Proof.} We discuss two cases. The first case is $d_{sr} \leq n_1-1$. Then the Singleton-like bound is $q^{n_1(n_1-d_{sr}+1)+n_2^2}$. The first part of ${\bf C}$ is consisting of $q^{n_1(n_1-d_{sr}+1)}$ codewords of the form $(a_0x+\cdots+a_{n_1-d_{sr}}x^{q^{n_1-d_{sr}}}, {\bf 0})$, where $a_0, a_1, \ldots, a_{n_1-d_{sr}}$ are $n-d_{sr}+1$ arbitrary elements on the field ${\bf F}_{q^{n_1}}$, and the $q$-polynomial is understood as a matrix in ${\bf F}_q^{(n_1, n_1)}$. The second matrix is the all-zero matrix. It is clear that the first part is a linear code with the minimum sum-rank distance at least $n_1-(n_1-d_{sr})=d_{sr}$.

The second part corresponds to $q^{n_2^2}$ codewords. First of all we decompose the ${\bf F}_q$-linear space ${\bf F}_{q^{n_1}} \cdot x^{q^{n_1-d_{sr}+1}}$ as the direct sum of $n_2$ linear subspaces ${\bf V}_1, \ldots, {\bf V}_{n_2}$,  of the dimension $\dim_{{\bf F}_q}({\bf V}_i)=n_2$, $i=1, \ldots, n_2$. This is guaranteed from the condition $n_1 \geq n_2^2$. Then each of $n_2$ subparts of the second part is a dimension $n_2$ code consisting of all codewords of the form $(a_{n_1-d_{sr}+1}x^{q^{n_1-d_{sr}+1}}, b_i x^{q^i})$, where $a_{n-d_{sr}+1} \in {\bf V}_i$, $b_i \in {\bf F}_{q^{n_2}}$, and moreover they are the same after a suitable ${\bf F}_q$ linear space isomorphism of ${\bf V}_i$ to ${\bf F}_{q^{n_2}}$. Therefore the dimension of each subpart is $n_2$. The second part is the direct sum of all these $n_2$ subparts and it is obvious the dimension is $n_2^2$. The minimum sum-rank distance is at least $n_1-(n_1-d_{sr}+1)+1=d_{sr}$ since $a_{n_1-d_{sr}+1}$ and $b_i$ are nonzero for a nonzero codeword.

When $d_{sr}=n_1+d$, $0 \leq d \leq n_2-1$, the Singleton-like bound is $q^{n_2(n_2-d+1)}$. The linear code ${\bf C}$ is the direct sum of $n_2-d+1$ subcodes of the dimension $n_2$. As in the first case, we decompose the ${\bf F}_q$-linear space ${\bf F}_{q^{n_1}} \cdot x^{q^0}$ as the direct sum of $n_2$ linear subspaces ${\bf V}_1, \ldots, {\bf V}_{n_2}$,  of the dimension $\dim_{{\bf F}_q}({\bf V}_i)=n_2$, $i=1, \ldots, n_2$. This is guaranteed from the condition $n_1 \geq n_2^2$. Then each of these $n_2-d+1$ linear subcodes is consisting of codewords of the form $(a_0^jx^{q^0}, b_j x^{q^{j-1}})$ where $a_0^j \in {\bf V}_j$ and $b_j \in {\bf F}_{q^{n_2}}$ for $j=1, \ldots, n_2-d+1$, and $a_0^j$ and $b_j$ are the same with a suitable linear isomorphism of ${\bf V}_j$ with ${\bf F}_{q^{n_2}}$. This is a dimension $n_2$ linear subcode. It is easy to verify that the minimum sum-rank distance is at least $n_1+n_2-(n_2-d)=n_1+d=d_{sr}$.

It is easy to verify that all these linear subcodes are linear independent. Hence the conclusion is proved.

\subsection{MSRD code construction}

In this section we give our explicit construction of linear MSRD codes of various square matrix sizes. This is a generalization of the construction of the block length two case in the previous subsection. From the following result it is clear that the constructed code attains the Singleton-like bound.

{\bf Theorem 5.2.} {\em Let $n_1>n_2>\cdots>n_t$ be $t$ positive integers. Let $d_{sr}=\Sigma_{i=1}^{j-1}n_i+d$ where $j \in \{1, \ldots, t\}$ and $0 \leq d \leq n_j-1$ be the unique representation of the minimum sum-rank distance. Suppose that $n_1, \ldots, n_t$ and $d_{sr}$ satisfy \\
1) $n_{j-1} \geq n_j(n_j-d+1)+n_{j+1}^2+\cdots+n_t^2$;\\
2) $n_j \geq n_{j+1}^2+\cdots+n_t^2$.\\
Then a linear MSRD code with $q^{n_j(n_j-d+1)+\Sigma_{i=j+1}^t n_i^2}$ codewords over an arbitrary field ${\bf F}_q$ can be constructed explicitly. The minimum sum-rank distance of this MSRD code is $d_{sr}$.}

{\bf Proof.} In each block position we have linearly independent ${\bf F}_q$-linear mappings $x^{q^{0}}, x^q, \ldots, x^{q^{n_i-1}}$ over ${\bf F}_{q^{n_i}}$ for $i=1, \ldots, t$.  In our construction, many copies of repetition codes over ${\bf F}_{q^{n_i}}$, $i=j, j+1, \ldots, t$, are used. When each such repetition code is used, new $x^{q^v}$'s in some block position are introduced, so the linearly independence is guaranteed. It is important that coefficients of some $x^{q^{v_i}}$ at different block positions are not zero for a nonzero codeword.

For each $x^{q^v}$ at the $i$-th block  position, the set of all coefficients is the field ${\bf F}_{q^{n_i}}$. This is a ${\bf F}_q$-linear space of the dimension $n_i$, therefore the linear space of all coefficients of $x^{q^v}$ at the $i$-th position, can be decomposed to the direct sum of $(n_j-d+1)+n_{j+1}+\cdots+n_t$ ${\bf F}_q$-linear subspaces ${\bf V}_w^i$, $i=1, 2, \ldots, j-1$ and $w=1, 2, \ldots, n_j-d+1+n_{j+1}+\cdots+n_t$, of dimensions $n_j, \ldots, n_j, n_{j+1}, \ldots, n_{j+1}, \ldots, n_t, \ldots, n_t$. If $i \leq j-1$, this is guaranteed from the condition $n_{j-1} \geq n_j(n_j-d+1)+n_{j+1}^2+\cdots+n_t^2$.

The dimension in the Singleton-like bound for sum-rank-metric codes is $n_j(n_j-d+1)+n_{j+1}^2+\cdots+n_t^2$. The first term $n_j(n_j-d+1)$ comes from $x^{q^0}, x^q, \ldots, x^{q^{n_j-d}}$ at the $j$-the block position. We use $n_j-d+1$ copies of length $j$ repetition code as in the proof of Theorem 2.1. Then for a nonzero codeword ${\bf c}=({\bf c}_1, \ldots, {\bf c}_j)$ in this repetition code, ${\bf c}_1, \ldots, {\bf c}_j$ are not zero. The dimension $n_j$ linear subspace of ${\bf F}_{q^{n_1}}$ of the coefficient of $x^{q^0}$ at the 1st block position is used for ${\bf c}_1$ with a suitable base of ${\bf F}_{q^{n_1}}$, ......, the dimension $n_j$ linear subspace of ${\bf F}_{q^{n_{j-1}}}$ of the coefficient of $x^{q^0}$ at the $(j-1)$-th block position is used for ${\bf c}_{j-1}$ with a suitable base of ${\bf F}_{q^{n_{j-1}}}$, at the $j$-th position, the coefficient of $x^{q^0}, \ldots, x^{q^{n_j-d}}$ at the $j$-th block position is used for ${\bf c}_j$, for the $j+1, \ldots, t$-th block positions, zero $q$-polynomials are used. The $q$-polynomials of these $(n_j-d+1)$ copies of codewords are as follows, $$M_i=\{({\bf c}_1)^i x^{q^0}, ({\bf c}_2)^i x^{q^0}, \ldots, ({\bf c_j})^ix^{q^i}, {\bf 0}, \ldots, {\bf 0})\},$$ where $i=0, 1, \ldots, n_j-d$, and $({\bf c_v})^i$'s, $v=1, \ldots, j$, are suitable images of ${\bf c}_v$'s in the ${\bf F}_{q^{n_v}}$ space of the coefficients. Then we have $q^{n_j(n_j-d+1)}$ codewords in the constructed sum-rank-metric code from these $n_j-d+1$ copies of the repetition code. The minimum sum-rank distance is at least $n_1+\cdots+n_{j-1}+(n_j-(n_j-d))=n_1+\cdots+n_{j-1}+d=d_{sr}$, since ${\bf c}_1, \ldots, {\bf c}_j$ are not zero for a nonzero codeword in the repetition codes.

When $i=1, \ldots, t-j$, for the $(i+1)$-th term $n_{j+i}^2$ in the dimension of the Singleton-like bound for sum-rank-metric codes, we use $n_{j+i}$ copies of the length $(j+1)$ repetition code. For a codeword ${\bf c}=({\bf c}_1, \ldots, {\bf c}_{j+1})$, it is obvious that ${\bf c}_1, \ldots, {\bf c}_{j+1}$ are not zero for a nonzero codeword. The dimension $n_{j+i}$ linear subspace of ${\bf F}_{q^{n_1}}$ of the coefficient of $x^{q^0}$ at the 1st block position is used for ${\bf c}_1$ with a suitable base of ${\bf F}_{q^{n_1}}$, ......, the dimension $n_{j+i}$ linear subspace of ${\bf F}_{q^{n_{j-1}}}$ of the coefficient of $x^{q^0}$ at the $(j-1)$-th block position is used for ${\bf c}_{j-1}$ with a suitable base of ${\bf F}_{q^{n_{j-1}}}$, at the $j$-th position, the dimension $n_{j+i}$ linear subspace of ${\bf F}_{q^{n_j}}$ of the coefficient of $x^{q^{n_j-d+1}}$ is used for ${\bf c}_j$ with a suitable base, since $n_j \geq n_{j+1}^2+\cdots+n_t^2$, at the $(j+1), \ldots, (j+i-1)$-th positions, only zero $q$-polynomials are used, the coefficient of $x^{q^0}, \ldots, x^{q^{n_{j+i}-1}}$ at the $(j+i)$-th block position is used for ${\bf c}_{j+i}$, for the $(j+i+1), \ldots, t$-th block positions,  zero $q$-polynomials are used. The $q$-polynomials of these $n_{j+i}$ copies of codewords are as follows,
\begin{align*}
  M_l=& \{({\bf c}_1)^l x^{q^0}, \ldots, ({\bf c}_{j-1})^lx^{q^0}, ({\bf c_j})^lx^{q^{n_j-d+1}}, {\bf 0}, \ldots, {\bf 0}, \\
  & ({\bf c}_{j+1})^lx^{q^l}, \ldots, {\bf 0})\},
\end{align*}
where $l=0, 1, \ldots, n_{j+i}$, $({\bf c_v})^i$'s, $v=1, \ldots, j$, are suitable images of ${\bf c}_v$'s in the ${\bf F}_{q^{n_v}}$ space of the coefficients, $({\bf c}_{j+1})^l$ is the suitable image in ${\bf F}_{q^{n_{j+i}}}$ space of coefficients. Then we have $q^{n_{j+1}^2}$ codewords in the constructed sum-rank-metric code from these $n_{j+1}$ copies of the repetition code. The minimum sum-rank distance is at least $n_1+\cdots+n_{j-1}+(n_j-(n_j-d+1))+1=n_1+\cdots+n_{j-1}+d=d_{sr}$, since ${\bf c}_1, \ldots, {\bf c}_{j+i}$ are not zero for a nonzero codeword in the repetition code.

The above constructed sum-rank-metric code is ${\bf F}_q$-linear. For two different codewords ${\bf x}_1$ and ${\bf x}_2$ in the above code, the difference ${\bf x}_1-{\bf x}_2$ has the rank at least $n_i$ at the $i$-th block position, for $i=1, \ldots, t-1$, and the sum-rank at least $d$,  at $j, j+1, \ldots t$-th block positions. Then the minimum sum-rank distance is at least $n_1+\cdots+n_{j-1}+d=d_{sr}$. The conclusion is proved.

{\bf Corollary 5.1.} {\em Let $n_1, n_2, \ldots, n_t$ be $t$ positive integers satisfying $n_i \geq n_{i+1}^2+\cdots+n_t^2$ for $i=1, 2, \ldots, t-1$. Then a linear MSRD code over an arbitrary finite field ${\bf F}_q$ with matrix sizes $n_1 \times n_1, \ldots, n_t \times n_t$, and any given minimum sum-rank distance can be constructed explicitly.}

{\bf Remark 5.1.} As in the two constructions of sum-rank-metric codes in Section II, the construction of linear MSRD codes in this section is a combination of the Hamming metric codes and the $q$-polynomial representation of rank-metric codes. The main point of the construction is as follows. From the condition $n_j\geq n_{j+1}^2+\cdots+n_t^2$, $j=1, \ldots, t-1$, there are sufficiently many $q$-polynomials with suitable degrees to construct a code attaining the Singleton-like bound.

\section{Conclusions}

In this paper three simple constructions of linear sum-rank-metric codes from the combination of Hamming metric codes and $q$-polynomial representations of rank-metric codes are proposed. Numerous good linear sum-rank-metric codes over ${\bf F}_q$, $q=2,3,4$, have been given. Many of these linear sum-rank-metric codes from the presently known best Hamming metric codes have larger dimensions when compared with previous codes of the same minimum sum-rank distances. Asymptotically good sum-rank-metric code sequences close to the Gilbert-Varshamov-like bound are also presented. These asymptotically good sequences of sum-eank-metric codes can be constructed by a polynomial-time algorithm. Explicit linear MSRD codes with square matrix sizes $n_1 \times n_1, \ldots, n_t \times n_t$ satisfying $n_i \geq n_{i+1}^2+\cdots+n_t^2$, $i=1, 2, \ldots, t-1$, over an arbitrary finite field,  are constructed for all possible minimum sum-rank distances. We show that the decoding of binary linear sum-rank-metric codes constructed in this paper can be reduced to the fast decodings in the  Hamming metric in our paper \cite{Chen}.

% or


\appendix
We list more small block size linear sum-rank-metric codes constructed from Theorem 2.1 and the presently known best Hamming metric codes of \cite{codetable}. These codes are compared with the codes constructed in \cite{MP21} and the Singleton-like bound. It is clear that most of our codes are larger than these previously constructed codes of the same sum-rank distances. Some of our codes are close to the Singleton-like bound.
\begin{table}[h!]
  \caption{\label{tab:A-q-5-3} Block length $t=15$, $n=m=2$.}
  \begin{center}
  \begin{tabular}{|l|l|l|l|}
  \hline
  $d_{sr}$&dimension &Table IV, \cite{MP21}&Singleton \\ \hline
  $4$& $2 \cdot 25$  & $2 \cdot 20$ & $2 \cdot 27$\\ \hline
  $5$ & $2 \cdot 21$  &$2 \cdot 18$ & $2 \cdot 26$ \\ \hline
  $6$ &$2 \cdot 20$  &$ 2 \cdot 16$ & $2 \cdot 25$ \\ \hline
  $7$ &$2 \cdot 18$ & $2 \cdot 14$ & $2 \cdot 24$ \\ \hline
  $8$ &$ 2\cdot 17$  & $2 \cdot 10$ & $2 \cdot 23$ \\ \hline
  $9$ &$ 2 \cdot 13$ & $2 \cdot 8$  & $2 \cdot 22$ \\ \hline
  $10$ &$2 \cdot 13$ &$2 \cdot 8$  & $2 \cdot 21$ \\ \hline
  $11$ &$2 \cdot 11$ &$ 2\cdot 6$  & $2 \cdot 20$ \\ \hline
  $12$ &$2 \cdot 10$ &$2 \cdot 4$  & $2 \cdot 19$ \\ \hline
  $13$ &$ 2 \cdot 8$ & none  & $2\cdot 18$ \\ \hline
  $14$ &$2 \cdot 8$ &$ 2\cdot 2$ & $2 \cdot 17$ \\ \hline
  $15$ &$2 \cdot 7$  &    none & $2 \cdot 16$  \\ \hline
  \end{tabular}
  \end{center}
  \end{table}

  Notice that from Theorem 2.1 and the presently known best linear codes over ${\bf F}_4$ in \cite{codetable}, our constructed linear sum-rank-metric codes have arbitrary block lengths $t \leq 256$. Therefore much more linear sum-rank-metric codes with relative good parameters can be obtained. Comparing with sum-rank-metric codes constructed in \cite{MP21}, our codes have more flexibilities of their parameters. For example the following table lists linear sum-rank-metric codes over ${\bf F}_2$ of the block size $17$ and the matrix size $2 \times 2$.

  \begin{table}[h!]
    \caption{\label{tab:A-q-5-3} Block length $t=17$, $n=m=2$.}
    \begin{center}
    \begin{tabular}{|l|l|l|}
    \hline
    $d_{sr}$&dimension &Singleton \\ \hline
    $4$& $2 \cdot 29$  & $2 \cdot 31$ \\ \hline
    $5$ & $2 \cdot 25$  &$2 \cdot 30$ \\ \hline
    $6$ &$2 \cdot 24$  &$2\cdot 29$ \\ \hline
    $7$ &$2 \cdot 22$ & $2 \cdot 28$\\ \hline
    $8$ &$ 2\cdot 21$  &$2 \cdot 27$ \\ \hline
    $9$ &$ 2 \cdot 20$ & $2 \cdot 26$\\ \hline
    $10$ &$2 \cdot 16$ &$2 \cdot 25$ \\ \hline
    $11$ &$2 \cdot 14$ &$2 \cdot 24$  \\ \hline
    $12$ &$2 \cdot 14$ &$ 2\cdot 23$    \\ \hline
    $13$ &$ 2 \cdot 11$ &$2\cdot 22$  \\ \hline
    $14$ &$2 \cdot 10$ &$2 \cdot 21$\\ \hline
    $15$ &$2 \cdot 9$  & $2 \cdot 20$  \\ \hline
    $16$ &$2 \cdot 8$  & $2 \cdot 19$  \\ \hline
    $17$ &$2 \cdot 7$  & $2 \cdot 18$  \\ \hline
    \end{tabular}
    \end{center}
    \end{table}

  % \begin{longtable}{|l|l|l|}
  % \caption{\label{tab:A-q-5-3} Block length $t=17$, $n=m=2$.}\\ \hline
  % $d_{sr}$&dimension &Singleton \\ \hline
  % $4$& $2 \cdot 29$  & $2 \cdot 31$ \\ \hline
  % $5$ & $2 \cdot 25$  &$2 \cdot 30$ \\ \hline
  % $6$ &$2 \cdot 24$  &$2\cdot 29$ \\ \hline
  % $7$ &$2 \cdot 22$ & $2 \cdot 28$\\ \hline
  % $8$ &$ 2\cdot 21$  &$2 \cdot 27$ \\ \hline
  % $9$ &$ 2 \cdot 20$ & $2 \cdot 26$\\ \hline
  % $10$ &$2 \cdot 16$ &$2 \cdot 25$ \\ \hline
  % $11$ &$2 \cdot 14$ &$2 \cdot 24$  \\ \hline
  % $12$ &$2 \cdot 14$ &$ 2\cdot 23$    \\ \hline
  % $13$ &$ 2 \cdot 11$ &$2\cdot 22$  \\ \hline
  % $14$ &$2 \cdot 10$ &$2 \cdot 21$\\ \hline
  % $15$ &$2 \cdot 9$  & $2 \cdot 20$  \\ \hline
  % $16$ &$2 \cdot 8$  & $2 \cdot 19$  \\ \hline
  % $17$ &$2 \cdot 7$  & $2 \cdot 18$  \\ \hline
  % \end{longtable}

  In the following table, the block length $t=63$ and the matrix size $2 \times 2$ binary linear sum-rank-metric codes are listed and compared with Table VI of \cite{MP21}. It is obvious that our codes are larger and closer to the Singleton-like bound.

  \begin{table}[h!]
  \caption{\label{tab:A-q-5-3} Block length $t=63$, $n=m=2$.}
  \begin{center}
    \begin{tabular}{|l|l|l|l|}
      \hline
      $d_{sr}$&dimension &Table VI, \cite{MP21}&Singleton \\ \hline
      $4$& $2 \cdot 119$  &$ 2 \cdot 112$ & $2 \cdot 123$ \\ \hline
      $5$ & $2 \cdot 114$  &$2 \cdot 108$ & $2 \cdot 122$ \\ \hline
      $6$ &$2 \cdot 112$  &$2 \cdot 106$ & $2 \cdot 121$ \\ \hline
      $7$ &$2 \cdot 107$ & $2 \cdot 100$ & $2 \cdot 120$ \\ \hline
      $8$ &$ 2\cdot 106$  & none & $2 \cdot 119$\\ \hline
      $9$ &$ 2 \cdot 101$ & none & $2 \cdot 118$\\ \hline
      $10$ &$2 \cdot 99$ &$2 \cdot 88$ & $2 \cdot 117$ \\ \hline
      $11$ &$2 \cdot 96$ &none  & $2 \cdot 116$ \\ \hline
      $12$ &$2 \cdot 95$ &none  & $2 \cdot 115$ \\ \hline
      $13$ &$ 2 \cdot 88$ & none  & $2 \cdot 114$\\ \hline
      $14$ &$2 \cdot 87$ &$2 \cdot 70$ & $2 \cdot 113$\\ \hline
      $15$ &$2 \cdot 84$  & none  & $2 \cdot 112$\\ \hline
      $16$ &$2 \cdot 82$  & none  & $2 \cdot 111$\\ \hline
      $17$ &$ 2\cdot 77$  & none & $2 \cdot 110$\\ \hline
      $18$ &$ 2 \cdot 77$ & none & $2 \cdot 109$\\ \hline
      $19$ &$2 \cdot 72$ &none & $2 \cdot 108$\\ \hline
      $20$ &$2 \cdot 71$ &none  & $2 \cdot 107$\\ \hline
      $21$ &$2 \cdot 70$ &none  & $2 \cdot 106$\\ \hline
      $22$ &$2 \cdot 69$ &$ 2\cdot 52$  & $2 \cdot 105$\\ \hline
      $23$ &$2 \cdot 67$&none & $2 \cdot 104$\\ \hline
      $24$ &$2 \cdot 66$&none & $2 \cdot 103$\\ \hline
      $25$ &$2 \cdot 58$&none & $2 \cdot 102$\\ \hline
      $26$ &$2 \cdot 58$&none & $2 \cdot 101$\\ \hline
      $27$ &$2 \cdot 56$&none & $2 \cdot 100$\\ \hline
      $28$ &$2 \cdot 55$&none  & $2 \cdot 99$\\ \hline
      $29$ &$2 \cdot 51$&none & $2 \cdot 98$\\ \hline
      $30$ &$2 \cdot 50$  &$ 2 \cdot 28$ & $2 \cdot 97$ \\ \hline
      $31$ &$2 \cdot 47$&none & $2 \cdot 96$\\ \hline
      $32$& $2 \cdot 46$  & none  & $2 \cdot 95$\\ \hline
      $38$ & $2 \cdot 37$  &$2 \cdot 16$ & $2 \cdot 89$\\ \hline
      $46$ &$2 \cdot 29$  & $2 \cdot 8$ & $2 \cdot 81$\\ \hline
      $54$ &$2 \cdot 20$ & $2 \cdot 8$ & $2 \cdot 73$\\ \hline
    \end{tabular}
  \end{center}
  \end{table}

  In the following table, the block length $t=127$ and the matrix size $2 \times 2$ linear binary sum-rank-metric codes are listed and compared with Table VII of \cite{MP21}. It is obvious that our codes are larger and closer to the Singleton-like bound.

  % \newpage

  \begin{table}[h!]
  \caption{\label{tab:A-q-5-3} Block length $t=127$, $n=m=2$.}
  \begin{center}
    \begin{tabular}{|l|l|l|l|}
      \hline
      $d_{sr}$&dimension &Table VII, \cite{MP21}&Singleton \\ \hline
      $4$& $2 \cdot 246$  &$ 2 \cdot 238$&$2 \cdot 251$ \\ \hline
      $5$ & $2 \cdot 240$  &$2 \cdot 233$ &$2 \cdot 250$\\ \hline
      $6$ &$2 \cdot 236$  &$2 \cdot 231$ &$2 \cdot 249$\\ \hline
      $7$ &$2 \cdot 231$ & $2 \cdot 224$&$2 \cdot 248$\\ \hline
      $10$ &$ 2\cdot 221$  &$2 \cdot 210$&$2 \cdot 245$ \\ \hline
      $14$ &$ 2 \cdot 202$ & $2 \cdot 189$ &$2 \cdot 241$ \\ \hline
      $22$ &$2 \cdot 172$ &$2 \cdot 154$&$2 \cdot 233$\\ \hline
      $30$ &$2 \cdot 146$ &$2 \cdot 112$&$2 \cdot 225$ \\ \hline
      $38$ &$2 \cdot 124$ &$2 \cdot 86$&$2 \cdot 217$    \\ \hline
      $46$ &$ 2 \cdot 106$ &$2 \cdot 70$&$2 \cdot 209$  \\ \hline
      $54$ &$2 \cdot 91$ &$2 \cdot 42$&$2 \cdot 201$ \\ \hline
      $62$ &$2 \cdot 77$  &$2 \cdot 28$&$2 \cdot 193$   \\ \hline
    \end{tabular}
  \end{center}
  \end{table}

  Numerous good small block length linear sum-rank-metric codes over ${\bf F}_3$ of the matrix size $n=m=2$ can be constructed from the presently known best small linear codes over ${\bf F}_9$ in \cite{codetable}. In the following tables , we give some such linear rank-sum-metric codes of the block length $t=31$. No previous code can be compared. Many of our codes are close to the Singleton-like bound.

  \begin{table}[h!]
  \caption{\label{tab:A-q-5-3} Block length $t=31$, $q=3$, $n=m=2$.}
  \begin{center}
    \begin{tabular}{|l||l|l|}
      \hline
      $d_{sr}$&dimension&Singleton \\ \hline
      $4$& $2 \cdot 57$&$2 \cdot 59$ \\ \hline
      $5$ & $2 \cdot 53$&$2 \cdot 58$  \\ \hline
      $6$ &$2 \cdot 52$&$2 \cdot 57$  \\ \hline
      $7$ &$2 \cdot 49$ &$2 \cdot 56$\\ \hline
      $8$ &$ 2\cdot 47$&$2 \cdot 55$  \\ \hline
      $9$ &$ 2 \cdot 44$&$2 \cdot 54$ \\ \hline
      $10$ &$2 \cdot 42$ &$2 \cdot 53$ \\ \hline
      $11$ &$2 \cdot 40$ &$2 \cdot 52$  \\ \hline
      $12$ &$2 \cdot 39$ &$2 \cdot 51$  \\ \hline
      $13$ &$ 2 \cdot 36$&$2 \cdot 50$   \\ \hline
      $14$ &$2 \cdot 35$ &$2 \cdot 49$\\ \hline
      $15$ &$2 \cdot 32$ &$2 \cdot 48$   \\ \hline
      $16$ &$2 \cdot 31$ &$2 \cdot 47$ \\ \hline
      $17$ &$2 \cdot 29$&$2 \cdot 46$    \\ \hline
      $18$& $2 \cdot 28$&$2 \cdot 45$  \\ \hline
      $19$ & $2 \cdot 25$&$2 \cdot 44$  \\ \hline
      $20$ &$2 \cdot 24$ &$2 \cdot 43$  \\ \hline
      $21$ &$2 \cdot 22$&$2 \cdot 42$ \\ \hline
      $22$ &$ 2\cdot 21$ &$2 \cdot 41$  \\ \hline
      $23$ &$ 2 \cdot 20$&$2 \cdot 40$  \\ \hline
      $24$ &$2 \cdot 19$ &$2 \cdot 39$ \\ \hline
      $25$ &$2 \cdot 18$   &$2 \cdot 38$ \\ \hline
      $26$ &$2 \cdot 17$   &$2 \cdot 37$  \\ \hline
      $27$ &$ 2 \cdot 15$  &$2 \cdot 36$\\ \hline
      $28$ &$2 \cdot 14$ &$2 \cdot 35$ \\ \hline
      $29$ &$2 \cdot 13$  &$2 \cdot 34$\\ \hline
      $30$ &$2 \cdot 13$ &$2 \cdot 33$  \\ \hline
    \end{tabular}
  \end{center}
  \end{table}

  Similarly many good small block length linear sum-rank-metric codes over ${\bf F}_4$ of the matrix size $n=m=2$ can be constructed from small length linear codes over ${\bf F}_8$, considered as linear codes over ${\bf F}_{16}$. In the following tables, we give some block length $21$ linear rank-sum-metric codes over ${\bf F}_4$ of the matrix size $2 \times 2$. Many of our codes are close to the Singleton-like bound.

  \begin{table}[h!]
  \caption{\label{tab:A-q-5-3} Block length $t=21$, $q=4$, $n=m=2$.}
  \begin{center}
    \begin{tabular}{|l|l|l|}
      \hline
      $d_{sr}$&dimension&Singleton  \\ \hline
      $4$& $2 \cdot 37$&$2 \cdot 39$  \\ \hline
      $5$ & $2 \cdot 33$&$2 \cdot 38$  \\ \hline
      $6$ &$2 \cdot 32$&$2 \cdot 37$  \\ \hline
      $7$ &$2 \cdot 30$&$2 \cdot 36$ \\ \hline
      $8$ &$ 2\cdot 29$&$2 \cdot 35$  \\ \hline
      $9$ &$ 2 \cdot 25$&$2\cdot 34$\\ \hline
      $10$ &$2 \cdot 24$&$2 \cdot 33$ \\ \hline
    \end{tabular}
  \end{center}
  \end{table}

  There have been few known linear binary sum-rank-metric codes of the matrix size $3 \times 3$ in the literature. We give some such codes from Theorem 2.1 and the presently known best linear codes over ${\bf F}_8$ in \cite{codetable}. Many of our constructed codes are close to the Singleton-like bound.

  \begin{table}[h!]
  \caption{\label{tab:A-q-5-3} Block length $t=12$, $q=2$, $n=m=3$.}
  \begin{center}
    \begin{tabular}{|l||l|l|}
      \hline
      $d_{sr}$&dimension&Singleton \\ \hline
      $4$& $3 \cdot 30$&$3 \cdot 33$ \\ \hline
      $5$ & $3 \cdot 27$&$3 \cdot 32$  \\ \hline
      $6$ &$3 \cdot 26$&$3 \cdot 31$  \\ \hline
      $7$ &$3 \cdot 22$ &$3 \cdot 30$\\ \hline
      $8$ &$3 \cdot 21$&$3 \cdot 29$  \\ \hline
      $9$ &$3 \cdot 19$&$2 \cdot 28$ \\ \hline
      $10$ &$3 \cdot 17$ &$3 \cdot 27$ \\ \hline
      $11$ &$3 \cdot 15$ &$3 \cdot 26$  \\ \hline
      $12$ &$3 \cdot 15$ &$3 \cdot 25$  \\ \hline
      $13$ &$3 \cdot 12$&$3 \cdot 24$   \\ \hline
      $14$ &$3 \cdot 12$ &$3 \cdot 23$\\ \hline
      $15$ &$3 \cdot 11$ &$3 \cdot 22$   \\ \hline
      $16$ &$3 \cdot 10$ &$3 \cdot 21$ \\ \hline
      $17$ &$3 \cdot 9$&$3 \cdot 20$    \\ \hline
      $18$& $3 \cdot 9$&$3 \cdot 19$  \\ \hline
      $19$ & $3 \cdot 7$&$3 \cdot 18$  \\ \hline
      $20$ &$3 \cdot 7$ &$3 \cdot 17$  \\ \hline
      $21$ &$3 \cdot 6$&$3 \cdot 16$ \\ \hline
      $22$ &$3 \cdot 5$ &$3 \cdot 15$  \\ \hline
      $23$ &$3 \cdot 5$&$3 \cdot 14$  \\ \hline
      $24$ &$3 \cdot 5$ &$3 \cdot 13$ \\ \hline
    \end{tabular}
  \end{center}
  \end{table}

  \begin{table}[h!]
  \caption{\label{tab:A-q-5-3} Block length $t=31$, $q=2$, $n=m=3$.}
  \begin{center}
    \begin{tabular}{|l||l|l|}
      \hline
      $d_{sr}$&dimension&Singleton \\ \hline
      $4$& $3 \cdot 89$&$3 \cdot 90$ \\ \hline
      $5$ & $3 \cdot 83$&$3 \cdot 89$  \\ \hline
      $6$ &$3 \cdot 82$&$3 \cdot 88$  \\ \hline
      $7$ &$3 \cdot 77$ &$3 \cdot 87$\\ \hline
      $8$ &$3 \cdot 75$&$3 \cdot 86$  \\ \hline
      $9$ &$3 \cdot 72$&$2 \cdot 85$ \\ \hline
      $10$ &$3 \cdot 70$ &$3 \cdot 84$ \\ \hline
      $11$ &$3 \cdot 67$ &$3 \cdot 83$  \\ \hline
      $12$ &$3 \cdot 66$ &$3 \cdot 82$  \\ \hline
      $13$ &$3 \cdot 61$&$3 \cdot 81$   \\ \hline
      $14$ &$3 \cdot 59$ &$3 \cdot 80$\\ \hline
      $15$ &$3 \cdot 56$ &$3 \cdot 79$   \\ \hline
      $16$ &$3 \cdot 54$ &$3 \cdot 78$ \\ \hline
      $17$ &$3 \cdot 52$&$3 \cdot 77$    \\ \hline
      $18$& $3 \cdot 51$&$3 \cdot 76$  \\ \hline
      $19$ & $3 \cdot 48$&$3 \cdot 75$  \\ \hline
      $20$ &$3 \cdot 47$ &$3 \cdot 74$  \\ \hline
      $21$ &$3 \cdot 44$&$3 \cdot 73$ \\ \hline
      $22$ &$3 \cdot 41$ &$3 \cdot 72$  \\ \hline
      $23$ &$3 \cdot 40$&$3 \cdot 71$  \\ \hline
      $24$ &$3 \cdot 39$ &$3 \cdot 70$ \\ \hline
    \end{tabular}
  \end{center}
  \end{table}

More binary linear sum-rank-metric codes from the presently known best codes in \cite{codetable} and Construction 2 are listed in \cite{Lao}. More sum-rank-metric codes from quaternary BCH codes, quaternary Goppa codes constructed in \cite{Chen} are also listed in \cite{Lao} for the convenience of readers. New explicit sum-rank-metric codes are welcomed to be included in the webpage \cite{Lao}.
% \appendices

% % you can choose not to have a title for an appendix
% % if you want by leaving the argument blank
% \section{}

\section*{Acknowledgment}
The author is grateful to three anonymous reviewers and the Associate Editor, Professor Camilla Hollanti, for their helpful comments and suggestions that improved the presentation of this paper.

\ifCLASSOPTIONcaptionsoff
  \newpage
\fi



%\bibliographystyle{IEEEtran}


\begin{IEEEbiographynophoto}{Hao Chen}
  obtained his Ph.D. degree in mathematics in the Institute of Mathematics, Fudan University in 1991. He is now a professor of the College of Information Science and Technology/Cyber Security, Jinan University. His research interests are coding and cryptography, quantum information and computation, lattices and  algebraic geometry.
\end{IEEEbiographynophoto}


%\newpage



%\vfill


\end{document}